# SYNTHETIC SOLAR X-RAY FLARES TIME SERIES SINCE 1968 AD

Boris Komitov[(1)], Peter Duchlev[(1)], Kaloyan Penev[(2)], Kostadinka Koleva[(1)], Momchil Dechev[(1)]

Institute of Astronomy, 72 Tsarigradsko Chaussee, Sofia 1784, Bulgaria
b_komitov@sz.inetg.bg

**Abstract.**
On the basis of the highest quality GOES satellites data for the period 1980-2009 AD multiple regression models of three detached "synthetic" series, consisting of the monthly numbers of soft X-ray flares (SXR) (separately for the classes C, M, X and total numbers) were built for the epoch 1968-2009 AD.

The monthly numbers of radiobursts at four frequencies (29-33, 609, 8800 and 15400 MHz ) and the radio flux at 2800 MHz (F10.7) are used as inputs. There are very strong relationships (correlation coefficients *R* are from 0.79 to 0.93 and Snedekor-Fisher's *F* parameter is in the range of 2.5-7). The earlier GOES X-ray data (1975-1979 AD) and the older SOLRAD satellite data (1968-1974 AD) are not precise enough. They suggest a probable decrease in SXR frequency due to the low instrumental sensitivity. It has also been found that the relatively weak sunspot Zurich cycle No 20 is significantly richer in the strongest SXR solar X flares than the next three cycles No 21-23.

Keywords: solar X-ray flares, solar radio emissions, sunspots, activity indices, solar cycle

*1. Introduction*

Solar flares and coronal mass ejections (CMEs) are among the most energetic and spectacular solar activity events, which can release a vast amount of plasma and magnetic flux into the outer space and cause interplanetary disturbances and geomagnetic storms near the Earth (Gosling. 1993; Webb *et al*., 1994). The physical relationship between flares and CMEs has been a long-standing elusive issue in solar physics (Kahler, 1992; Gosling, 1993; Hundhausen, 1999). Nevertheless, recent studies demonstrate that there is a strong physical connection between flares and CMEs (Cheng *et al*., 2010a, b, and references therein). They proceed on a different time and spatial scales, and can occur in association, but are independent of each other too. Both observations and theories of these eruptive phenomena suggest that they are often involved in the reorganization of the same large scale magnetic fields in the corona (e.g. Shibata, 1999; Priest and Forbes, 2002). However it has to be noted that the time evolution of different solar activity parameters, does not match exactly due to the complicated dynamics active in the different levels of the solar atmosphere (Kane, 2006).

Practically the solar flares emit all kinds of wavelengths (from gamma to radio waves), which originate in different regions of the solar atmosphere. Many of the flares are the primary sources of radiobursts (e.g. Kim *et al*. 2009; Lobzin *et al*. 2010), as well as a significant increase of the solar electromagnetic radiation in the X-ray and EUV ranges (e.g. Sterling et al. 2000; Tripathi et al. 2004). The strongest flares (e.g. SXR class X) - often associated with interplanetary CMEs are related to the so called "Forbush-decreases" (FD), i.e. the temporary fading of galactic cosmic rays fluxes from 3 to 15-20% in comparison with their common levels. Such strong flares are also sources of solar energetic particles (SEP), such as protons, electrons or heavy ions with energies up to 1 GeV or higher (Reames, 2004). The SEP penetration into the middle and lower Earth atmosphere is very high. There are few tens of cases when radiation enhancement events penetrated to the ground level between February 28[th], 1942 and now (May, 2010) that were caused by SEP.

For this reasons, the strong solar flares and the associated with them CMEs, SEP-events and geomagnetic storms are of high importance for many processes in the middle and low Earth atmosphere. Moreover, these events play a crucial role for the stability and security of many components of the human technological infrastructure

(communications, electric power grids, electronic equipments, radiative environments of the upper troposphere aviation, etc.).

Reliable predictions of the solar flare activity are an important task both from the scientific and applied points of view. Especially the predictions of the powerful X-ray flares of classes M and X need good algorithms. However, the correlation between strong solar flare classes and the sunspot activity (Schwabe-Wolf's cycles) on long time scales is relatively low (see Table 1). Killic (2009) and Kane (2009) have been pointed out that unlike to the radio index F10.7, the differences between solar flares and sunspots behavior are very significant during the decreasing phase of the Zurich cycle 23. According to the analysis of Kane (2009), the last statement could be generalized for all corpuscular events in the solar atmosphere and interplanetary space. In this respect, it should be noted that a significant part of the most powerful solar flares of class X have occurred between AD 2003 and AD 2007, during the fading solar cycle 23, i.e. during middle and low sunspot activity conditions. For the purposes of predicting solar flare activity, one should build models based not only on the relationships between flares and sunspot activity, but also on other solar events, whose physical nature is more closely related to those one of the flares. Such events could be the radiobursts of different types for example. These models cannot be used directly to make predictions, because they require knowledge of the preliminary the levels of the predictors, on which the flare activity depends. However they could help to detect some important dynamical features of the flares, such as trends and statistically important cycles. Achieving this goal requires as long and as homogeneous data series as possible.

In AD 1936 (Link and Kleczek, 1949), the epoch of relatively regular solar flare observations in the optical range began. But we note that some uncertainty concerning the data during the World War II exists. There is a homogeneous series since January 1976 that is published and regularly updated by the National Geophysical Data Center (Knoška, and Petrásek, 1984; Ataç and Özgüç, 1998). The regular observations of the solar flares in the soft X-ray (SXR) spectral range by GOES satellite series are provided since September 1975. SXR flares are classified according to the order of magnitude of the peak burst intensity measured in 1-8 Å band as follows: C-class ($10^{-6} \leq I < 10^{-5}$ W/m$^2$), M-class ($10^{-5} \leq I < 10^{-4}$ W/m$^2$), X-class ($I \geq 10^{-4}$ W/m$^2$). There also exist earlier data from the SOLRAD satellite from March 1968 up to February 1974. Thus an empty "window" without observations and duration of approximately 1.5 years remains.

To the present day (May 18, 2010), the observational data of GOES SXR flares cover almost 35 years. If the GOES data could be connected to the older one of the SOLRAD satellite the total period of SXR flares data will cover 42 years. Such an extension could help with searching for temporary tendencies (cycles and trends) longer than one Schwabe-Wolf's cycle. This statement refers to overall SXR flare events, as well as the separated strong flare classes C, M and X, and their relative distributions during the four sequential solar cycles.

However there could be a serious problem with directly using the older data before AD 1980 in a time series analysis or any other numerical studies. The precision of the data presented in the National Geophysical Data Center GOES X-ray after 1980 AD has significantly increased. After this year a new, weaker than class C SXR flare - class B ($10^{-7} \leq I < 10^{-6}$ W/m$^2$) was added to the data sets. This is an argument for considering the GOES data as being of higher quality and more precise after the 1980 AD period of time. Taking into account this fact, one could assume that the SXR flares of C-class are not properly classified in the data sets before 1980 AD.

The main aim of this study is to build synthetic series for the monthly numbers of SXR solar flares of C, M and X classes since January 1968. Our study is predominantly based on the highest quality part of GOES satellite data (AD 1980-2009). On the basis of SXR flares data sets the multiple regression models for the monthly numbers of C, M and X-classes flares are obtained. As factors (predictors) the monthly number of the radiobursts at four frequencies in the MHz and GHz ranges and the standard solar index F10.7 (radio flux at 2800 MHz) are used. There are two reasons for using radiobursts monthly numbers: 1) The relationships between the radiobursts and solar flares events (e.g. Shanmugaraju *et al.* 2003, Huang, Yan, and Liu, 2008), 2). In some stations the long-term radiobursts data sets have good coverage for the period since 1965-1967 AD. Thus, they could be used for successful reconstruction of the solar SXR flares since the beginning of the AD 1968 period on the basis of the obtained multiple models. In this work for comparison with the synthetic data from the corresponding periods, the earlier SOLRAD and GOES satellite data (before AD 1980) are also used.

*2. Data and methods*

The solar flares are strong non-stationary processes with a duration in the range of minutes. Therefore they should be studied against not only the relative smoothly changing indices, like the daily values of the International sunspot number *Ri*, or the radio-index *F10.7*, but also with other non-stationary phenomena, like the radiobursts in the MHz and GHz spectral ranges. On the other hand the flares influence the solar wind parameters and consequently the galactic cosmic ray (GCR) flux. Thus, the data for the GCR neutron fluxes (NFs), measured in Moscow station and used as a proxy for the solar wind, and their relationships with the flares of C, M and X-classes has been studied. All data necessary for this study, SXR flares (SOLRAD and GOES data), International sunspot numbers *Ri*, solar radiobursts, *F10.7* radio-index, and GCR NF, are published and regularly updated in National Geophysical Data Center (NGDC) at the web link ftp://ftp.ngdc.noaa.gov/STP/SOLAR_DATA.

On the basis of the SXR solar flares and radiobursts data, and also by using our own primary data proceeding software -the daily, monthly, and yearly numbers of these phenomena with selected characteristics have been calculated and organized as time series.

As a first step of this study we searched for pairwise correlation relations between the monthly numbers of SXR flares of C, M and X classes and the similar indices of *Ri*, *F10.7* and GCR *NF* for the period January 1980-December 2009. For the next step, we used a multiple correlation-regression analysis, in which the monthly numbers of radiobursts at different frequencies in MHz and GHz ranges were added to better express the relations. The comparison between the separate multi-factor models and the selection of the best ones among them was made on the basis of the multiple coefficients of correlation and the Snedekoor-Fisher's *F*-parameter. Finally, we use the obtained multiple models to extrapolate backwards in time the monthly numbers of C, M and X flare classes and to create their "synthetic" data series since January 1968.

*3. Results and analysis*

*3.1. The soft X-ray solar flares, sunspots, F10.7 index, and cosmic rays*

The pairwise coefficients of linear correlation *r* between the monthly values of international sunspot numbers (*Ri*), solar radio flux at $\lambda$=10.7 cm (*F10.7*), the neutron flux data from Moscow station (*NF*), and the monthly numbers of solar flare classes C ($N_C$), M ($N_M$) and X ($N_X$) for the period AD 1980-2009 are listed in Table 1.

Table 1. Pairwise coefficients of linear correlation *r* between the monthly values of 6 solar and solar-modulated indices for the period January 1980 - December 2009

|       | Ri     | F10.7  | NF     | $N_C$  | $N_M$  | $N_X$  |
|-------|--------|--------|--------|--------|--------|--------|
| Ri    | 1      | 0.975  | -0.798 | 0.847  | 0.719  | 0.449  |
| F10.7 | 0.975  | 1      | -0.796 | 0.856  | 0.759  | 0.487  |
| NF    | -0.798 | -0.796 | 1      | -0.724 | -0.615 | -0.464 |
| $N_C$ | 0.847  | 0.856  | -0.724 | 1      | 0.672  | 0.415  |
| $N_M$ | 0.719  | 0.759  | -0.615 | 0.672  | 1      | 0.770  |
| $N_X$ | 0.449  | 0.487  | -0.464 | 0.415  | 0.770  | 1      |

All coefficients of correlation listed in Table 1 are statistically significant. The negative values of *r* for the *NF* data are as a result of the Forbush-effect, i.e. the reverse relationship between the GCR (*NF)* and the overall solar activity level. Clearly, the $N_C$ index (C-class flares) is significantly more tightly related to the *Ri* and *F10.7* than the $N_M$ and $N_X$ indices of the stronger M- and X-flares. The corresponding values of *r* for $N_C$ in relation to *Ri* and *F10.7* are >0.8, but the relationship with *F10.7* is slightly better than the one with *Ri*. One must bare in mind that the correlation between *F10.7* and *Ri* is very high during the studied period (*r* =+0.975).

In accordance with the last two facts one can conclude that the high value of *r* between *Ri* and $N_C$ is rather due to the very closed relationship between *Ri* and *F10.7* . Thus, the real and important correlation is between *F10.7* and $N_C$ and there is no significant independent influence of *Ri* on $N_C$.

The correlations of the $N_M$ and $N_X$ indices with *Ri* and *F10.7* are significantly lower, but like $N_C$ their relation to *F10.7* is stronger than the one to *Ri* (Table 1). This is in contrast with the very low correlation ($0.4 < |r| < 0.5$) of the X-class flares ($N_X$) to almost all another indices in Table 1, except that of the M-class flares ($N_M$) for which *r* =+ 0.77. Such low values of *r* could be explained by the specifics of the X-class definition (see Section 1). The criterion for classifying a SXR flare to X-class is that its intensity exceeds $10^{-4}$ W/m$^2$ . There is no upper limit for X-class flares, unlike the flare classes C and M. So, to the X-class could be comprised of events with very different energy characteristics, which probably occurred in very different physical conditions in the solar atmosphere. Usually the X-class covers many flares, which are up to 8-10 times stronger than M-class flares. Moreover, there are some relatively rare and extremely strong SXR flares called "mega–flares", such as the events on October, 28th and November, 4th, 2003 ("Halloween storms"), the big flare in March 13[th], 1989 etc. The so-called "Carrington flare" in September 2[nd], 1859 most probably belongs to such extreme phenomena (Cliver, 2006). The energy of these flares exceeds by a factor of $10^6$ or even more that of the X-class events close to the lower limit of the X-class intensity ($10^{-4}$ W.m$^{-2}$). This non-homogeneity leads to a very high variance within the X-flare class and as a result of that to the low *r* values in Table 1.

The absolute values of *r* in Table 1 between *NF* on one hand and $N_C$, $N_M$ and $N_X$ on the other are significantly weaker than the ones with *Ri* and *F10.7*. Such correlations are caused by the fact that the flares, especially the strongest ones of M and X-class, are primary sources of interplanetary coronal mass ejections (ICME's) and therefore lead to Forbush-decrease of the GCR-flux (Kane, 2008). This should change the coefficient *r* between *NF* and the monthly numbers of flares to less negative, i.e. to smaller absolute values.

At this stage of the study the result suggest that the most important and independent predictor of the three flare classes is the index F10.7. It is used in the multiple analyses during the next step.

*3.2. The X-ray flares and radiobursts (1980-2009 AD ): multiple regression models*

The best correlations of the monthly solar flare numbers are to the F10.7 index. However, as it is shown in Table 1 only the value of *r*= +0.856 for the C-class flares seems high enough to consider the relationship useful for reliable predictions and extrapolations. For M-class flares (*r* = +0.759), searching for a better multiple statistical relation is advisable, while for X-class flares the correlation (*r* = +0.487) is so weak that the building and using of multiple statistical models is inevitable.

It is customary to search for relationships between the solar flares and other solar fast non-stationary events such as CME's or radiobursts. In our opinion, the radioburst data are the most suitable for the purposes of this stage of our study. There are continuous data since the middle of the 1960s published in NGDC for radiobursts from many stations at different frequencies. After a large number of statistical numerical experiments with this data, it has been found that the better predictors for the SXR flares are the radiobursts from the following data sets (Table 2).

Table 2. The used radiobursts data sets (1968 – 2009 AD)

| Frequency [MHz] | Station |
|---|---|
| 29- 33 | Upice (Czhech Republic) * |
| 606- 609 | Sagamore Hill ( MA, USA) |
| 8800 | Sagamore Hill (MA, USA) |
| 15400 | Sagamore Hill  (MA,USA) |

* no data before May 1972 are available

In contrast to radiobursts, the regular observations of CMEs started in 1996 AD as a part of the Solar and Heliospheric Observatory (SOHO) satellite program. There are also CME data for two separated periods in the 80s ( 1980 and 1984-1989 AD ) from the Solar Maximum Mission (SMM) satellite. There are no data during, the interesting for our study, solar cycle No 20 (SC 20). Regular optical flares data in the NGDC are available since January 1976 comprising a relatively short, continuous data series.

Two of the radioburst frequencies are given in the ranges 29-33 MHz and 606-609 MHz. This is because of the changes during the studied period from one observed frequency to another, situated very close to the first one. In these cases, we consider that there are no significant differences and the observed radiobursts are related to the same phenomena.

Three basic criteria are crucial for our choice of radioburst frequencies. 1) The period of observations should be long enough and it must cover the whole time interval since 1968 AD.  2) The used frequency set should be optimally covered in the MHz and GHz ranges, meaning that the monthly number of every used radioburst frequency is a factor (predictor) with enough detectable statistical weight for the SXR flares monthly numbers. 3) The radiobursts monthly numbers are not closely related to the overall solar activity indices such as F10.7. Because the correlation coefficients of the four selected radiobursts frequencies span the range 0.4-0.65 (Table 3, they could be taken as sufficiently independent factors in the multiple regression models. In that case it is necessary to take into account that the radiobursts at different frequencies are related to physical conditions in different layers of the solar atmosphere during non-stationary processes like flares.

As it is shown in Table 3 the correlation coefficients between  the monthly numbers of radiobursts at $f=$ 606, 8800 and 15400 MHz are high ($r \geq 0.89$), i.e. they are quite closely related to each other. The reason for using the selected three radioburst types is that including them simultaneously in our multiple regression models slightly improves them and on the other hand the Snedecor Fisher's F-parameter increases on order of 2% to 5%.

Table 3. The pair coefficients of linear correlation between the monthly radiobursts numbers and F10.7 index

|  | $F107$ | $N_{29}$ | $N_{606}$ | $N_{8800}$ | $N_{15400}$ |
|---|---|---|---|---|---|
| $F107$ | 1 | 0.485 | 0.622 | 0.586 | 0.529 |
| $N_{29}$ | 0.485 | 1 | 0.532 | 0.484 | 0.420 |
| $N_{606}$ | 0.622 | 0.532 | 1 | 0.912 | 0.893 |
| $N_{8800}$ | 0.586 | 0.484 | 0.912 | 1 | 0.963 |
| $N_{15400}$ | 0.529 | 0.420 | 0.893 | 0.963 | 1 |

$F_{10.7}$ – the solar radio flux at $\lambda=10.7$ cm (2800 MHz) in units $10^{-21}$ W.m$^{-2}$ Hz$^{-1}$

$N_{29}$ , $N_{606}$, $N_{8800}$  and $N_{15400}$ - the monthly numbers of radiobursts at 29-33, 606-609 , 8800 and 15400 MHz

After a lot of statistical experiments we found that for the epoch January 1980 – December 2009 the following multiple minimized functions are the best approximations for the considered C, M, and X classes of SXR solar flares.

a) C-class flares:

$$N_C = -222.57 + 0.6667 \cdot N_{29} + 0.6685 \cdot N_{8800} + 3.77 \, F_{10.7} - 0.00859 \cdot F_{10.7}^2 \quad (1)$$

$R = 0.894; \; F = 4.92$

b) M-class flares:

$$N_M = 14.65 - 0.0124 \cdot N_{29} - 0.0094 \cdot N_{606} - 0.5733 \cdot N_{8800} + 0.3213 \cdot N_{15400} - 0.362 \, F_{10.7} + 0.001723 \cdot F_{10.7}^2 \quad (2)$$

$R = 0.927; \; F = 6.95$

c) X-class flares:

$$N_X = 2.06 - 0.0093 \cdot N_{29} + 0.01034 \cdot N_{606} + 0.0169 \cdot N_{8800} + 0.111 \cdot N_{15400} - 0.0381 \, F_{10.7} + 0.000155 \cdot F_{10.7}^2 \quad (3)$$

$R = 0.791; \; F = 2.63$

The meaning of the symbols used in Equations (1-3) are as follows:
$N_C$, $N_M$ and $N_X$ – the monthly numbers of C, M and X class SXR flares;
$F_{10.7}$ – the solar radio flux at $\lambda=10.7$ cm (2800 MHz) in units $10^{-21}$ W.m$^{-2}$Hz$^{-1}$;
$N_{29}$, $N_{606}$, $N_{8800}$ and $N_{15400}$ - the monthly numbers of radiobursts at 29-33, 606-609, 8800 and 15400 MHz;
$R$ – the coefficient of multiple correlation ($0 \leq R \leq 1$);
$F$ – the Snedekor-Fisher's parameter ($F = S_t^2/S_0^2$, where $S_t^2$ is total variance and $S_0^2$ is residual variance ).

The monthly number of C-class flares in Equation (1) increases with the increase of the monthly radiobursts numbers at 29 and 8800 MHz, while statistically significant independent relations to the other two monitored frequencies (606 and 15400 MHz) are absent. On the other hand, the number of C-class events increases with the F10.7 in the whole real range of changes of this index. However the presence of the non-linear term $F_{107}^2$ indicates that there is a slight reversing effect at the very high levels of F107.

Unlike C-class flares the relationships between M and X flare classes and the radiobursts are much more complicated (Equations (2) and (3)). The M-class flares monthly numbers are decreased with the increasing of monthly numbers of radiobursts at 29, 606 and 8800 MHz, while their changes with the radiobursts at 15400 MHz are positive. The X-class flares monthly numbers are decreased only with the increasing of the radiobursts at 29 MHz, while their changes relatively to the other radioburst types are positive.

The negative linear terms including F10.7 index in the both Equations (2) and (3) indicate anticorrelation of the radio flux with the M- and X-class flares at low and medium levels of this index. The relationship is changed to positive when F10.7 > 210 for the M-class or F10.7 >245 for X-class flares. The C-class monthly numbers increase slowly at higher levels of F10.7, which is marked by the negative sign of the quadratic term in Equation (1). One could conclude that the increase of the M- and X-class flare activity is most probably due to the reduction of the C-class events. Furthermore, such reducing of the C-class flare activity could probably lead to the shifting of the mean energy of flare activity from lower to higher energetic classes. Thus it could explain also why the critical "threshold" for X-class flares (F10.7=245) is higher than M-class flares (F10.7=210).

It is interesting to compare the models (Equations 1-3) with the "pure" $F_{10.7}$ – models for the flare monthly numbers $N_C$, $N_M$ and $N_X$, i.e. if the radiobursts are fully excluded as factors. We have investigated relationships of the form $N_V = a.F_{10.7}^2 + b.F_{107} + c$ (V is C, M or X). In this case the corresponding R and F values decrease: $R = 0.884$ and $F = 4.53$ for $N_C$, $R = 0.782$ and $F = 2.56$ for $N_M$, and $R = 0.512$ and $F = 1.35$ for $N_X$. These results demonstrate that including the radiobursts plays a significant role for the C-class flare models and are very important for the models of M-class and especially of X-class flares. The minimal critical values of F at 95% and 99% statistical significance are 1.16 and 1.19, respectively.

### 3.3. Synthetic SXR flares data series (1968-2009 AD)

Equations (1-3) were obtained on the base of GOES data from the last 30 years (1980-2009 AD). We can consider them as the most representative and accurate for the flare classes C, M and X. We use these equations for a reconstruction model of the monthly numbers of SXR flare classes for the earlier period (1968-1979 AD). In our opinion, the older satellite data are not accurate enough, especially concerning the weaker C-class flares (see Section 3.4) and for this reason we prefer to model data for this older period.

The radioburst observations in Sagamore Hill at the above mentioned frequencies (Table 2) are continuous since 1967 AD, but the radiobursts observations at ~ 29 MHz in Upice are only available after May 1972. In view of that, we use Equations (1-3) to obtain the synthetic data for the period 1973-1979. For the earlier period 1968-1972 AD additional multiple models were obtained. They are only based on the indices $F_{10.7}$, $N_{606}$, $N_{8800}$ and $N_{15400}$. The obtained multiple regression equations are as follows:

$$N_C = -236.6 - 0.247.N_{606} + 0.986.N_{8800} + 4.097.F_{10.7} - 0.00953.F_{10.7}^2 \quad (4)$$

$R = 0.890; \ F = 4.76$

$$N_M = 14.86 - 0.015..N_{606} + 0.567.N_{8800} + 0.33.N_{15400} - 0.321 F_{10.7} + 0.00174.F_{10.7}^2 \quad (5)$$

$R = 0.923; \ F = 6.94$

$$N_X = 2.21 + 0.0061.N_{606} + 0.0124.N_{8800} + 0.1178.N_{15400} - 0.0418 F_{10.7} + 0.000166.F_{10.7}^2 \quad (6)$$

$R = 0.790; \ F = 2.62$

The comparison of the two groups of models (Equations 1-3) and (Equations 4-6) suggests that the decrease of R and F is more significant only for the C-class flares when $N_{29}$ is excluded (Equation 4). The F-parameter in Equation (4) for C-class model decreases by about 3.3% in comparison with those in Equation (1). This fact suggests that the monthly numbers of radiobursts at 29 MHz has almost the same relative participation in the total variance of $N_C$. For the M and X flare classes the excluding of $N_{29}$ leads to a very small effect ($\leq 0.4\%$). Thus, it does not significantly matter what model we will use: 2 or 5 for M-class and 3 or 6 for the X-class.

The built synthetic time series for the monthly numbers of C, M and X classes of SXR flares are plotted on Figures 1, 2 and 3. The corresponding tabulated data are given in Appendix A, B and C.

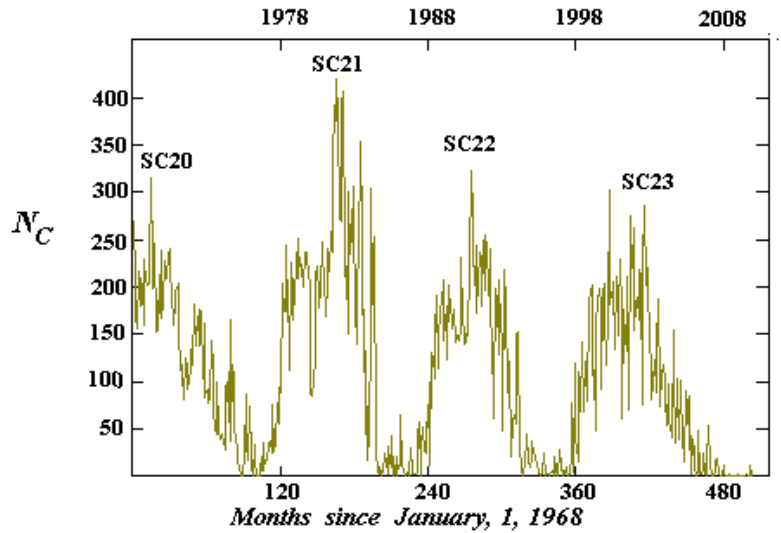
**Fig.1. Synthetic series of monthly numbers of C-class SXR flares (1968-2009 AD)**

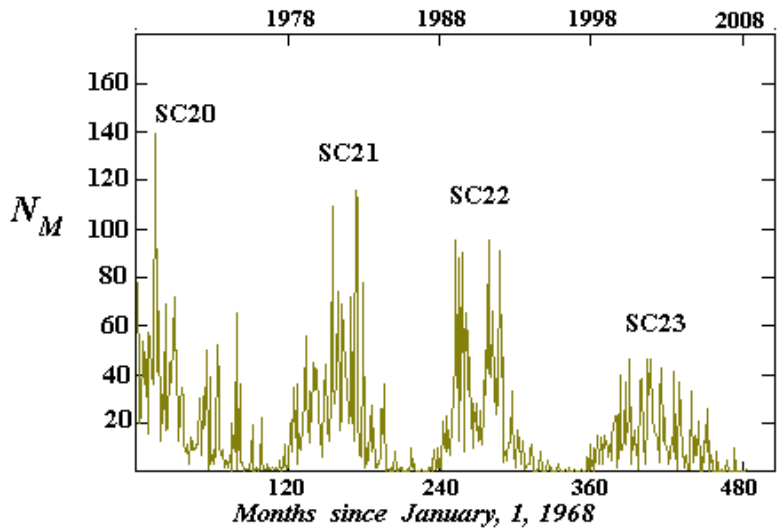
**Fig.2. Synthetic series of monthly numbers of M-class SXR flares (1968-2009 AD)**

The comparison of the plotted data on Figure 1, 2, and 3 suggests two essential phenomena:

**1.** The overall flare activity during the Zurich SC 20 is comparable with the next two cycles (SC 21 and SC 22). This feature is more visible in the stronger flare classes M and X than in class C. Such flare activity is in contrast to the sunspot activity rate during SC 20, which is known as the weakest one among the last four cycles SC 20-23.

**2.** During the last four decades a clearly visible downward trend in flare activity is present. This long-term tendency is better visible in the activity of flare classes M and X. Hence, the flare activity during the sunspot SC 23 is weaker than in the previous three cycles.

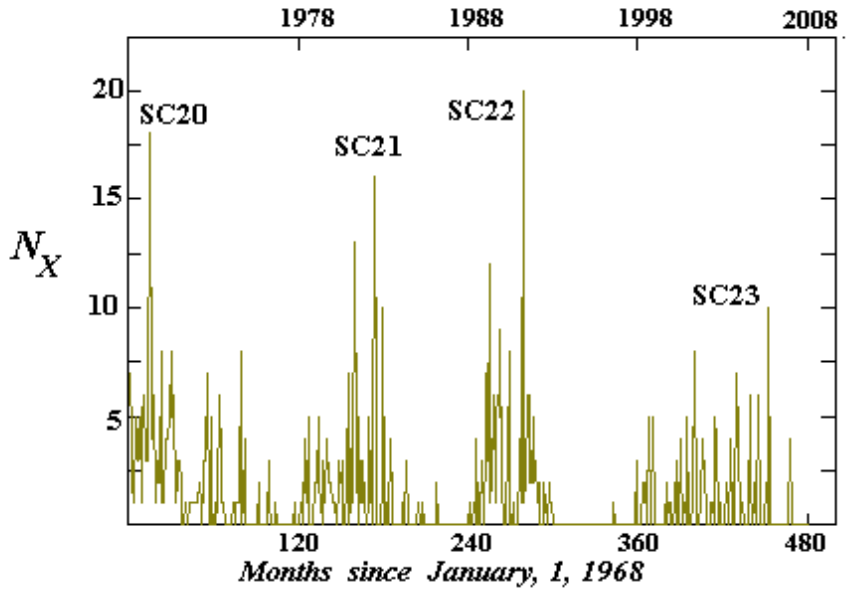

**Fig.3. Synthetic series of monthly numbers of X-class SXR flares (1968-2009 AD )**

*3.4. Synthetic and satellite solar X-ray flares data before 1980 AD*

The obtained synthetic data for the period before 1980 AD was compared with the earlier satellite data provided since March 1968. By using a least squares procedure, a number of linear relationships between the monthly numbers of C, M and X flare classes of the observed and synthetic data were determined. This procedure was performed separately for the period of SOLRAD observations (1968-1974 AD) and the early one of GOES observations (1975-1979 AD). As it was mentioned above there are no satellite observations between March 1974 and August 1975. The coefficients of linear correlations $r$ and the corresponding $F$-parameters were also determined. The results are shown in Table 4.

Table 4. The linear relationships ($N_{synth} = A.N_{obs}+B$) between the synthetic and the satellites data series for SXR flares in the epoch before 1980 AD

| Satellite | Flare class | A | B | r | F |
|---|---|---|---|---|---|
| SOLRAD | C | 1.79 | 91.44 | 0.837 | 3.30 |
| SOLRAD | M | 1.11 | 7.75 | 0.761 | 2.35 |
| SOLRAD | X | 0.75 | 1.19 | 0.689 | 1.87 |
| GOES | C | 2.00 | 11.78 | 0.919 | 6.26 |
| GOES | M | 0.975 | 3.58 | 0.852 | 3.57 |
| GOES | X | 0.63 | 0.22 | 0.780 | 2.50 |

The first important conclusion that can be made by examining the results in Table 4 is that there are significant differences between the synthetic and satellite data in the earliest part of the series, i.e. in the epoch of SOLRAD satellite observations before March 1974. As it is shown the coefficients of correlation $r$ for the three strong flare classes and the corresponding $F$-values of the SOLRAD data are significantly lower than the $r$ and $F$ values of the earlier GOES data (AD 1975-1979). Most probably, these differences are due to the higher general uncertainty of the SOLRAD data in comparison with the GOES data. The offsets of the regression formulas ($B$-parameters) are significantly higher for the SOLRAD data than for the GOES data. On the other hand, for both the SOLRAD and the GOES data the *B*-terms are higher for the C-class flares than for the M and X classes. These results suggest a relatively low sensitivity of the earlier satellite instruments compared to those after 1980 AD. Therefore, the earlier GOES and especially the SOLRAD instruments have not detected a significant part of SXR flares in the least energetic C-class. An additional argument for this conclusion is the fact that the coefficients *A* for C-class in both cases are in the range of 1.8-2, which indicates that the reported monthly numbers of C-class events before 1980 AD are less than the real ones if they are only estimated on the basis of satellite observations. The synthetic and satellite data for SXR flares of M-class before 1980 AD have the best coincidence. This can be seen by noting that $A \approx 1$ and the *B*-coefficients are relatively low for both series.

The relationships between the observed and synthetic X-class flares are different. The coefficients *A* are less than 1 ( 0.75 for SOLRAD and 0.63 for GOES). However, the B-terms are positive (0.22 for GOES and 1.19 for SOLRAD). Thus the differences between the synthetic and observed data are small and negligible for small monthly numbers $N_X$, but they can reach up to 30-40% during the months with high X-class flares activity when $N_X \geq 10$. At this stage of the study it is difficult to explain the causes for those differences and therefore they will be the subject of a future study.

### 3.5. "Epignosis"- test

For testing of the validity and stability of the extrapolation procedure of models (1-3) before 1980 one additional test has been made. It could be qualified as "epignosis", because it is very similar to the classical procedures of comparison between the predicted on the base of models and the real dynamics of the parameters being studied, which is commonly used in weather predictions.

In our case the monthly C, M and X-class flare numbers for the period January 1990 – December 2009 were used with the same factors as in models (1-3) for building multiple regression models. It is important to note, that the best minimized functions are of the same kinds as the general models for 1980-2009 AD. The corresponding multiple coefficients of correlation *R* between the original data and the obtained models and *F* –parameters are as follow: $R = 0.937$ and $F = 8.10$ for C-class; $R = 0.934$ and $F = 7.66$ for M-class ; $R = 0.791$ and $F = 2.63$ for X-class.

As it is shown by comparison of SXR flare values with the corresponding parameters of the general relationships (formulas 2 and 3) for M and X –classes, they remain practically the same or are slightly better if only the recent parts, after 1989 AD are taken when building the models. The difference is more significant for the C-class models. The *R* and *F* values for the general data series model (1) are noticeably weaker if it is compared to the model for the recent epoch (1990 – 2009 AD). The corresponding general model *F*- parameter is 4.92 , i.e there is about 1.64 times higher variance in comparison with the latter part (after 1989 AD), where $F = 8.10$.

We use the models, based on the data after 1989 AD for an extrapolation (prediction) in the past, during the period 1980-1989 AD. As it has been already noted, this "epignosis - test" should be considered as critical for the validity and reliability of the obtained synthetic time series in their early parts (1968-1989 AD). The modeled vs. real data are shown in Figure 4.

The extrapolation in the past is obviously very successful for the M and X-class flares. Both models (the red lines) caught very well all significant local maxima in the activity in these SXR classes, both their times of occurrence and relatively well their magnitudes.

It is very interesting to compare the model and real data for the C-class flares (the upper panel of Figure 4). It is clearly shown that the coincidence is very good in the middle and recent parts of the period after 1983 AD (approximately the 48$^{th}$ month). The differences are significant in the earliest part, between 1980 and 1983 AD, i.e. in the near maximum phase of SC 21.  The model data are significantly lower than the real ones between 1981 and 1983 AD (~30-40%) and higher during the first half of 1980 AD. What causes these differences is not easily explained at this stage.  Some of them could be caused by instrumental effects, for example during the first half of 1980 AD. On the other hand the significant underestimation by the model of the very high C-class activity during the period 1981-1983 AD needs additional analysis.

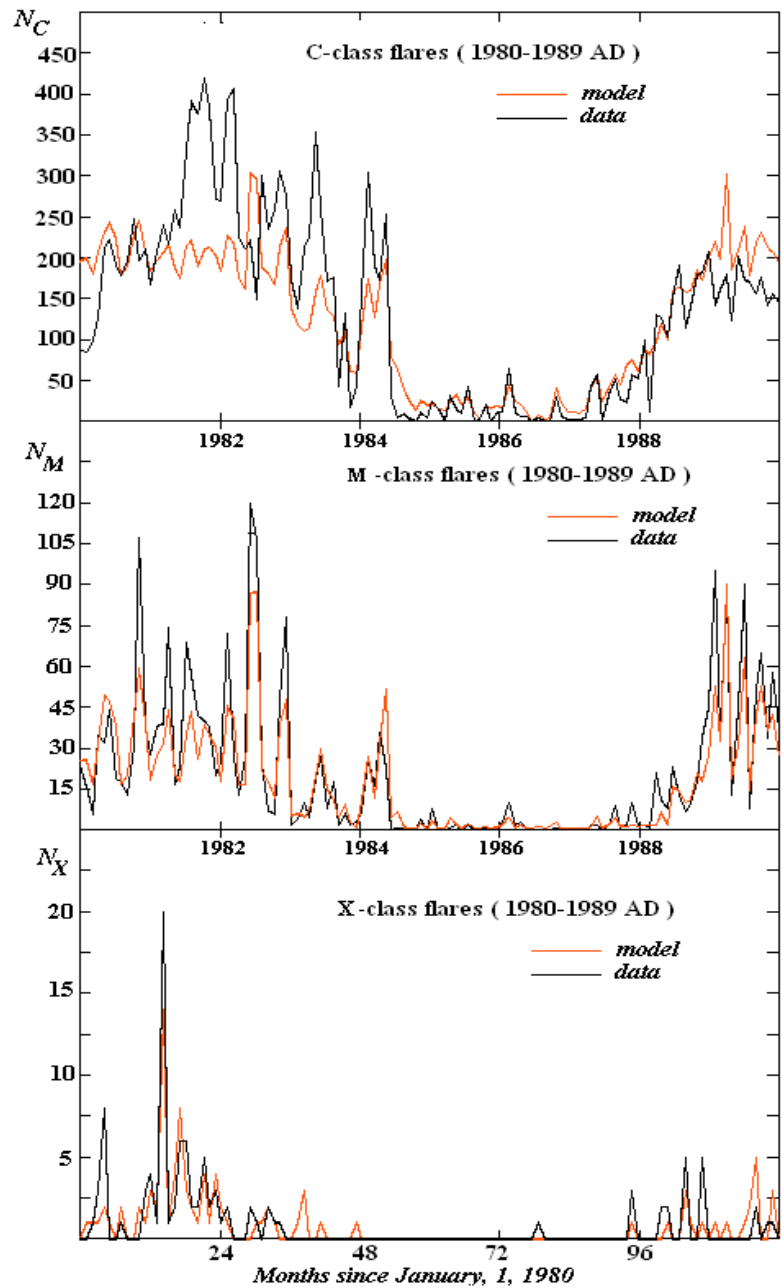

Fig.4. The models and instrumental data of the SXR flares during the period 1980-1989 AD.

**4.** *Discussion*

In our opinion, the proposed in this study synthetic time series of the monthly numbers of SXR flares of C, M and X classes is more realistic in their earlier (modeled) parts (before 1980 AD) than the direct instrumental data from SOLRAD and earlier GOES satellite observations. Our time series models for the epoch 1968-1979 AD are based on multiple regression models with high statistical significance. The data for F10.7 and the monthly numbers of radiobursts at four frequencies for the period 1980-2009 AD are used as input parameters. On the basis of this, we use the GOES satellite SXR flares data for the same period with two assumptions: 1) These data are more certain than the earlier one, before 1980 AD; 2) There are no significant physical causes for less validity of the obtained regression models during the earlier epochs.

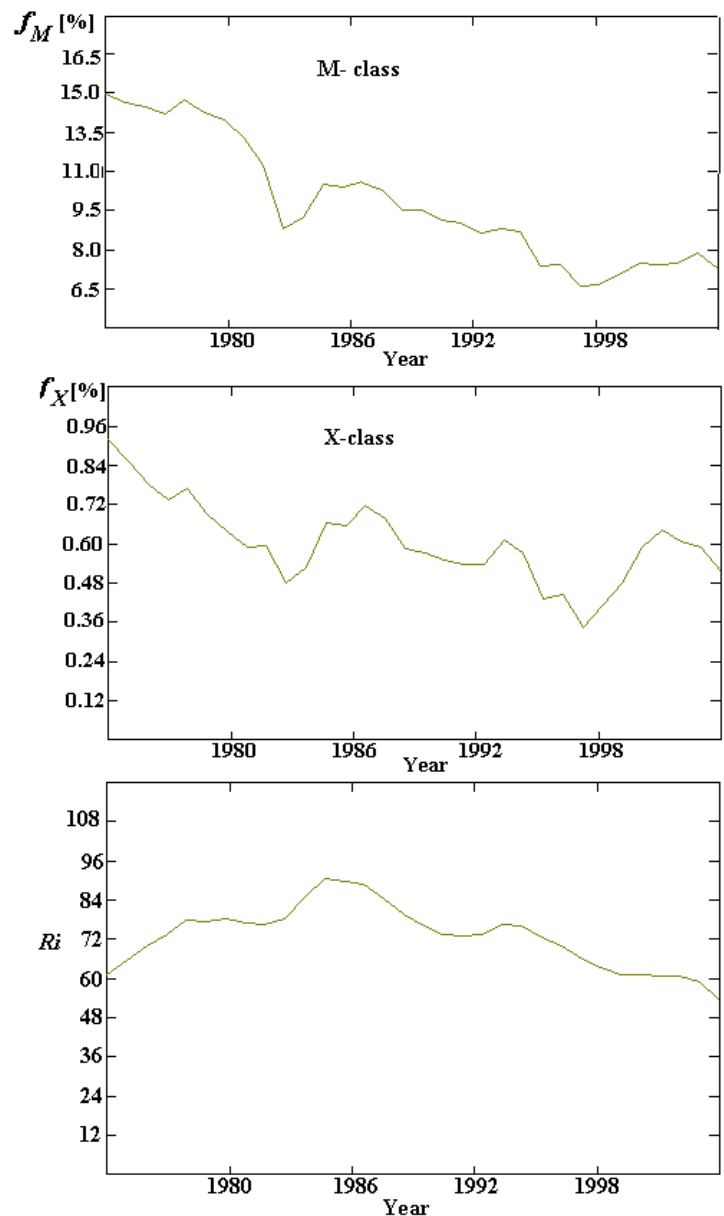

Fig. 5 Yearly relative parts of M- and X-class flares during the period AD 1968-2009. The panel of the bottom shows the corresponding 11-year smoothed Ri-values.

As it is shown on Figure 1, Figure 2, and Figure 3 the Zurich SC 20, which is the weakest in sunspot activity from the last four (SC 20-23) is comparable in SXR flare activity to the next three cycles (SC 21-23). This fact is even more evident when comparing the monthly numbers of M and X class flares in SC 20 to those of the next three cycles. The downward phase of SC 20 is very rich of M- and X-flares, even in the final pre-minimal phase before the start of SC 21 (Figure 2 and Figure 3), while there is total absence of M and X class flares during next epochs of the sunspot minima in 1986, 1996 and 2007-2009 AD. Every one of these free from M- and X-flares time windows is wider than the previous one. A low activity of C-class flares is shown in 1986 and 1996 AD, and during the last deep minimum between SC 23 and 24 it is very close to zero (Figure 1). Figure 2 and Figure 3 show that SC 23 is significantly poorer in M and X class flares than the previous three cycles. The maximum of the X-class flare activity during this cycle is strongly delayed (3-4 years) in relation to the sunspot maximum. This delay is mainly due to the relatively high X-class flares activity in the period of the Halloween storms (October-November 2003) and the next 3 years before the deep solar minimum between SC 23 and 24.

All these facts suggest the existence of a well expressed downward trend from sunspot SC 20 to SC 23 for the activity of strong SXR flares of classes M and X, and the extreme (minimum) phase is reached near to or a little before the maximum of SC 23. To elucidate such effect, we use the smoothed 11 year values of SXR flare numbers $N_C$, $N_M$, and $N_X$ to calculate the ratios $f_M = N_M / N_{tot}$ and $f_X = N_X/N_{tot}$, where $N_{tot}= N_C+N_M+N_X$. We took $f_M$ and $f_X$ equal to zero in all cases when $N_{tot} = 0$. Thus $f_M$ and $f_X$ present the relative parts (frequencies) of M and X class flares in the overall flare activity (C+M+X flares) for the corresponding time. The smoothed over 11 years values of $f_M$ and $f_X$ are shown in percent in Figure 5.

During SC 20 the relative parts of M and X-class flares is higher than during the next three sunspot cycles. A clear downward tendency is shown for the relative presence of the both M and X class flares since the maximum of SC 20. It seems that this tendency reaches its deepest point in the second half of the 1990s, near 1998 AD. The high M and X class flare activity during 2003-2006 AD is obviously the main factor that stops this downward trend. Unfortunately our synthetic time series are too short to allow us to make some more concrete assumptions and predictions for the M and X class flare activity in the near future and especially for SC 24. It is very possible that the negative trend in the behavior of the relative parts M and X class flares (Figure 4) is just a decreasing phase of a cycle with subcentury or quasi-century duration. If it is so, an increase in the relative frequencies of the strong flares of M and X classes could occur even during the obviously low in sunspots SC 24. However, it is still difficult to take or reject this assumption at this stage.

However, there are two important questions: 1. In what direction could the possible differences be between the calculated "synthetic" flare activity data and the measured ones before 1980 AD?; 2. How large are they? ; 3. How could these differences affect our conclusions, concerning the general downward tendency of M and X-flares activity during the whole investigated period?

As it is shown in Figure 5 ( "epignosis"- test) the largest absolute differences between the two data series during the period 1980-1989 AD occur during the most powerful local peaks – up to 30-40%. As it has been already pointed out in Section 3.5 the model caught very well the moments of main flare activity peaks, but the measured data values are usually larger than the modeled ones in these cases. This is consistent with the type of regression used for our models, which are based on Legendre's principle. On the other hand, in the middle and lower range of values all C, M and X-flares model data are in very good agreement with the instrumental ones.

On the basis of this, one could assume as most probable, that during the moments of largest peaks of SC20 and the upward phase of SC 21, the real level of SXR flare activity could been even slightly larger than the "synthetic" ones. This conclusion is related to the all three flare classes. Thus, there are no arguments, that the downward trend of the relative parts $f_M$ and $f_X$ should be significantly changed from what is presented in Figure 5. All conclusions concerning this part of the study remain the same.

The presented here "synthetic" series could and should be subjected to further attempts to improve it. In the first place, one should assume, that there are many radiobursts in the used frequency ranges which have not been detected in Sagamore Hill or Upice for different reasons. This is why, for a better representation of the radioburst activity during the last decades, data from other stations should alse be used as additional predictors of the flare activity. But there is a problem – the generated data series from the other stations are shorter as in both examples mentioned above. Another possibility is to try and use data for the optical flares - the continuous series after 1976 AD as well as older ones. Our preliminary analisys over the monthly data during the period 1980-2008 AD indicates that the relationships between SXR and optical flares (the Kleczek's index) are not very close ( $r$ = +0.78, +0.86 and +0.71 for C,M and X- class correspondingly).

*5. Concusions*

The main results obtained in this study could be summarized in four points:
1. On the base of the GOES satellite instrumental data for the solar SXR flares during the period 1980-2009 AD multiple regression models for the monthly numbers of C, M and X-class flares were built. In these models the mean monthly radio index F10.7 and the monthly numbers of radioburst at four frequencies ~29 MHz, ~606 MHz, 8800 MHz, and 15400 MHz are used as input parameters. The models are characterized by coefficients of multiple correlations in the range 0.78-0.93. The corresponding residual variance of these models is in the range of 14% (for class M) to 40% (for class X).
2. Using the regression models, synthetic time series of the monthly numbers of C, M and X-class SXR flares were built for the period 1968-2009 AD.
3. Evidence for the existence of a general downward trend of the SXR flare activity during the last decades of the 20$^{th}$ century was found. The trend is very well expressed in the yearly relative frequencies of the strong flare classes M and X. Moreover, this negative trend reached its deepest minimum during the increasing phase of SC 23. Additional evidence for the validity of this conclusion on the basis of an "epignose-test" has been found.
4. Most probably the rate of C-class SXR flare activity before 1980 AD has been underestimated because of a lack of instrumental sensitivity during the earlier satellite observations with SOLRAD and GOES.

Acknowledgements

The authors are thankful to National Geophysical Data Center, Boulder, Colorado, U.S.A. for proving the data of SXR flares (SOLRAD and GOES data), International sunspot numbers, solar radiobursts, *F10.7* radio-index, and GCR neutron flux via [ftp://ftp.ngdc.noaa.gov/STP](ftp://ftp.ngdc.noaa.gov/STP).

APPENDIX A: Monthly and Yearly Numbers of C-Class Flares (AD 1968-2009)

| Year | Jan. | Feb. | Mar. | Apr. | May | Jun. | Jul. | Aug. | Sep. | Oct. | Nov. | Dec. | Total |
|------|------|------|------|------|-----|------|------|------|------|------|------|------|-------|
| 1968 | 269 | 210 | 169 | 156 | 217 | 205 | 180 | 203 | 160 | 230 | 207 | 201 | 2407 |
| 1969 | 203 | 203 | 328 | 213 | 198 | 245 | 153 | 185 | 155 | 187 | 238 | 167 | 2475 |
| 1970 | 184 | 228 | 210 | 210 | 234 | 241 | 214 | 195 | 160 | 183 | 194 | 198 | 2451 |
| 1971 | 204 | 154 | 112 | 117 | 108 | 81 | 125 | 117 | 93 | 102 | 114 | 134 | 1461 |
| 1972 | 118 | 182 | 138 | 141 | 164 | 176 | 137 | 174 | 104 | 161 | 83 | 90 | 1668 |
| 1973 | 93 | 77 | 82 | 143 | 115 | 70 | 50 | 44 | 100 | 50 | 39 | 44 | 907 |
| 1974 | 41 | 40 | 29 | 92 | 99 | 58 | 166 | 37 | 118 | 106 | 57 | 36 | 879 |
| 1975 | 33 | 15 | 11 | 1 | 2 | 0 | 23 | 87 | 34 | 18 | 74 | 20 | 318 |
| 1976 | 11 | 1 | 34 | 19 | 1 | 1 | 0 | 13 | 7 | 36 | 7 | 17 | 147 |
| 1977 | 18 | 32 | 33 | 38 | 24 | 75 | 27 | 36 | 77 | 65 | 56 | 132 | 613 |
| 1978 | 89 | 203 | 175 | 173 | 243 | 175 | 150 | 112 | 225 | 196 | 165 | 195 | 2101 |
| 1979 | 221 | 252 | 209 | 231 | 219 | 220 | 200 | 237 | 236 | 230 | 218 | 198 | 2671 |
| 1980 | 87 | 85 | 100 | 127 | 213 | 222 | 191 | 178 | 197 | 248 | 197 | 211 | 2056 |
| 1981 | 168 | 208 | 240 | 217 | 259 | 237 | 332 | 391 | 376 | 420 | 383 | 271 | 3502 |
| 1982 | 270 | 394 | 406 | 225 | 211 | 222 | 150 | 300 | 235 | 259 | 306 | 277 | 3255 |
| 1983 | 174 | 139 | 213 | 226 | 353 | 260 | 171 | 176 | 44 | 132 | 17 | 43 | 1948 |
| 1984 | 157 | 305 | 190 | 173 | 253 | 30 | 5 | 9 | 4 | 0 | 11 | 5 | 1142 |
| 1985 | 25 | 16 | 2 | 31 | 15 | 10 | 42 | 1 | 3 | 20 | 1 | 11 | 177 |
| 1986 | 12 | 64 | 11 | 7 | 6 | 0 | 5 | 0 | 1 | 30 | 6 | 2 | 144 |
| 1987 | 3 | 2 | 5 | 44 | 57 | 1 | 33 | 52 | 27 | 24 | 57 | 52 | 357 |
| 1988 | 99 | 12 | 131 | 127 | 103 | 152 | 191 | 114 | 141 | 179 | 182 | 208 | 1639 |
| 1989 | 143 | 161 | 179 | 123 | 202 | 174 | 170 | 157 | 176 | 142 | 157 | 145 | 1929 |
| 1990 | 143 | 154 | 231 | 168 | 151 | 139 | 142 | 160 | 149 | 224 | 323 | 278 | 2262 |
| 1991 | 240 | 201 | 244 | 172 | 206 | 236 | 231 | 179 | 246 | 254 | 197 | 247 | 2653 |
| 1992 | 191 | 241 | 160 | 157 | 61 | 87 | 198 | 188 | 129 | 207 | 160 | 143 | 1922 |
| 1993 | 49 | 218 | 183 | 89 | 110 | 121 | 23 | 21 | 58 | 64 | 55 | 151 | 1142 |
| 1994 | 153 | 21 | 34 | 6 | 1 | 7 | 9 | 45 | 17 | 16 | 9 | 18 | 336 |
| 1995 | 37 | 27 | 21 | 14 | 16 | 2 | 1 | 2 | 2 | 24 | 2 | 0 | 148 |
| 1996 | 2 | 0 | 1 | 4 | 6 | 1 | 20 | 8 | 0 | 0 | 28 | 11 | 81 |
| 1997 | 0 | 6 | 2 | 9 | 6 | 1 | 3 | 21 | 77 | 5 | 119 | 37 | 286 |
| 1998 | 33 | 16 | 110 | 69 | 141 | 68 | 68 | 122 | 103 | 80 | 184 | 194 | 1188 |
| 1999 | 201 | 107 | 113 | 48 | 161 | 183 | 199 | 182 | 92 | 193 | 203 | 172 | 1854 |
| 2000 | 118 | 171 | 303 | 181 | 195 | 193 | 205 | 135 | 211 | 148 | 174 | 229 | 2263 |
| 2001 | 174 | 61 | 176 | 145 | 132 | 211 | 71 | 275 | 211 | 228 | 262 | 155 | 2101 |
| 2002 | 182 | 189 | 166 | 218 | 212 | 75 | 225 | 286 | 201 | 220 | 202 | 147 | 2323 |
| 2003 | 112 | 82 | 98 | 124 | 89 | 136 | 187 | 87 | 73 | 106 | 118 | 104 | 1316 |
| 2004 | 83 | 53 | 97 | 56 | 53 | 41 | 154 | 111 | 35 | 86 | 103 | 40 | 912 |
| 2005 | 102 | 30 | 42 | 16 | 91 | 54 | 81 | 29 | 85 | 3 | 39 | 27 | 599 |
| 2006 | 10 | 0 | 6 | 48 | 2 | 3 | 7 | 25 | 5 | 0 | 15 | 53 | 174 |
| 2007 | 12 | 3 | 0 | 1 | 6 | 17 | 19 | 3 | 0 | 0 | 0 | 12 | 73 |
| 2008 | 3 | 0 | 0 | 2 | 0 | 0 | 0 | 0 | 0 | 0 | 2 | 1 | 8 |
| 2009 | 0 | 0 | 0 | 0 | 0 | 0 | 2 | 0 | 1 | 11 | 0 | 14 | 28 |

**APPENDIX B:  Monthly and Yearly Numbers of M-Class Flares (AD 1968-2009)**

| Year | Jan. | Feb. | Mar. | Apr. | May | Jun. | Jul. | Aug. | Sep. | Oct. | Nov. | Dec. | Total |
|---|---|---|---|---|---|---|---|---|---|---|---|---|---|
| 1968 | 78 | 37 | 20 | 23 | 54 | 46 | 36 | 46 | 16 | 57 | 55 | 38 | 506 |
| 1969 | 43 | 36 | 139 | 45 | 41 | 66 | 13 | 32 | 20 | 30 | 69 | 17 | 551 |
| 1970 | 18 | 45 | 45 | 37 | 55 | 72 | 50 | 50 | 13 | 24 | 28 | 35 | 472 |
| 1971 | 30 | 12 | 8 | 8 | 11 | 3 | 14 | 13 | 9 | 9 | 11 | 10 | 138 |
| 1972 | 11 | 30 | 12 | 7 | 27 | 35 | 19 | 50 | 1 | 39 | 3 | 5 | 239 |
| 1973 | 9 | 3 | 8 | 52 | 36 | 11 | 7 | 6 | 9 | 2 | 5 | 2 | 150 |
| 1974 | 2 | 7 | 0 | 20 | 11 | 8 | 65 | 1 | 16 | 36 | 4 | 4 | 174 |
| 1975 | 0 | 1 | 1 | 1 | 1 | 1 | 6 | 19 | 0 | 0 | 2 | 0 | 32 |
| 1976 | 1 | 1 | 22 | 5 | 1 | 1 | 1 | 3 | 1 | 1 | 1 | 2 | 40 |
| 1977 | 1 | 2 | 1 | 1 | 2 | 5 | 0 | 2 | 11 | 4 | 1 | 4 | 34 |
| 1978 | 2 | 20 | 8 | 35 | 27 | 14 | 36 | 3 | 16 | 13 | 9 | 19 | 202 |
| 1979 | 30 | 56 | 30 | 21 | 12 | 33 | 21 | 45 | 34 | 43 | 42 | 26 | 393 |
| 1980 | 24 | 16 | 6 | 35 | 32 | 44 | 19 | 18 | 13 | 31 | 109 | 38 | 385 |
| 1981 | 28 | 38 | 39 | 74 | 17 | 23 | 69 | 57 | 42 | 40 | 38 | 20 | 485 |
| 1982 | 25 | 72 | 27 | 13 | 27 | 120 | 106 | 25 | 7 | 6 | 48 | 78 | 554 |
| 1983 | 2 | 4 | 10 | 5 | 19 | 27 | 8 | 18 | 2 | 6 | 2 | 3 | 106 |
| 1984 | 7 | 26 | 16 | 36 | 23 | 0 | 1 | 1 | 0 | 0 | 4 | 0 | 114 |
| 1985 | 8 | 0 | 0 | 1 | 2 | 0 | 2 | 0 | 0 | 1 | 0 | 0 | 14 |
| 1986 | 3 | 10 | 2 | 3 | 1 | 0 | 0 | 0 | 0 | 1 | 0 | 0 | 20 |
| 1987 | 0 | 0 | 0 | 2 | 2 | 0 | 2 | 9 | 1 | 2 | 10 | 2 | 30 |
| 1988 | 2 | 1 | 21 | 11 | 8 | 23 | 13 | 7 | 10 | 18 | 34 | 45 | 193 |
| 1989 | 95 | 35 | 88 | 13 | 45 | 90 | 8 | 52 | 65 | 34 | 58 | 37 | 620 |
| 1990 | 25 | 10 | 28 | 28 | 21 | 28 | 13 | 25 | 16 | 11 | 25 | 50 | 280 |
| 1991 | 32 | 52 | 103 | 41 | 39 | 66 | 29 | 33 | 24 | 53 | 27 | 91 | 590 |
| 1992 | 39 | 47 | 4 | 8 | 5 | 7 | 12 | 12 | 33 | 24 | 7 | 4 | 202 |
| 1993 | 2 | 17 | 13 | 3 | 5 | 13 | 4 | 1 | 2 | 3 | 3 | 8 | 74 |
| 1994 | 11 | 2 | 0 | 0 | 0 | 1 | 1 | 8 | 0 | 1 | 0 | 1 | 25 |
| 1995 | 0 | 5 | 1 | 2 | 0 | 0 | 0 | 0 | 0 | 3 | 0 | 0 | 11 |
| 1996 | 0 | 0 | 0 | 1 | 0 | 0 | 2 | 0 | 0 | 0 | 1 | 0 | 4 |
| 1997 | 0 | 0 | 0 | 1 | 1 | 0 | 0 | 1 | 6 | 0 | 11 | 1 | 21 |
| 1998 | 5 | 0 | 10 | 4 | 15 | 4 | 3 | 14 | 9 | 3 | 15 | 12 | 94 |
| 1999 | 10 | 6 | 11 | 5 | 16 | 17 | 23 | 23 | 2 | 8 | 40 | 9 | 170 |
| 2000 | 9 | 14 | 37 | 11 | 20 | 21 | 51 | 3 | 14 | 11 | 17 | 7 | 215 |
| 2001 | 10 | 1 | 37 | 38 | 11 | 13 | 3 | 22 | 50 | 32 | 46 | 47 | 310 |
| 2002 | 22 | 18 | 15 | 16 | 14 | 4 | 30 | 43 | 13 | 22 | 12 | 10 | 219 |
| 2003 | 8 | 3 | 7 | 14 | 8 | 41 | 7 | 4 | 1 | 37 | 24 | 6 | 160 |
| 2004 | 12 | 2 | 5 | 7 | 1 | 1 | 33 | 26 | 4 | 9 | 15 | 7 | 122 |
| 2005 | 21 | 1 | 0 | 0 | 13 | 4 | 19 | 8 | 26 | 0 | 7 | 4 | 103 |
| 2006 | 0 | 0 | 0 | 8 | 0 | 0 | 1 | 0 | 0 | 0 | 0 | 5 | 14 |
| 2007 | 0 | 0 | 0 | 0 | 0 | 10 | 0 | 0 | 0 | 0 | 0 | 0 | 10 |
| 2008 | 0 | 0 | 1 | 0 | 0 | 0 | 0 | 0 | 0 | 0 | 0 | 0 | 1 |
| 2009 | 0 | 0 | 0 | 0 | 0 | 0 | 0 | 0 | 0 | 0 | 0 | 0 | 0 |

APPENDIX C:  Monthly and Yearly Numbers of X-Class Flares (AD 1968-2009)

| Year | Jan. | Feb. | Mar. | Apr. | May | Jun. | Jul. | Aug. | Sep. | Oct. | Nov. | Dec. | Total |
|------|------|------|------|------|-----|------|------|------|------|------|------|------|-------|
| 1968 | 7 | 4 | 2 | 1 | 5 | 4 | 3 | 5 | 1 | 5 | 6 | 3 | 46 |
| 1969 | 4 | 3 | 18 | 4 | 4 | 6 | 1 | 3 | 2 | 2 | 8 | 1 | 56 |
| 1970 | 1 | 4 | 4 | 4 | 5 | 8 | 5 | 6 | 1 | 2 | 3 | 3 | 46 |
| 1971 | 2 | 0 | 0 | 1 | 1 | 0 | 1 | 1 | 1 | 1 | 1 | 1 | 10 |
| 1972 | 1 | 2 | 1 | 0 | 2 | 4 | 3 | 7 | 0 | 5 | 0 | 0 | 25 |
| 1973 | 1 | 0 | 1 | 6 | 3 | 1 | 1 | 0 | 0 | 0 | 0 | 0 | 13 |
| 1974 | 0 | 1 | 0 | 1 | 1 | 1 | 8 | 0 | 1 | 4 | 0 | 0 | 17 |
| 1975 | 0 | 0 | 0 | 0 | 0 | 0 | 0 | 2 | 0 | 0 | 0 | 0 | 2 |
| 1976 | 0 | 0 | 3 | 1 | 0 | 0 | 0 | 1 | 0 | 0 | 0 | 0 | 5 |
| 1977 | 0 | 0 | 0 | 0 | 0 | 0 | 0 | 0 | 1 | 0 | 0 | 0 | 1 |
| 1978 | 0 | 1 | 0 | 4 | 2 | 1 | 5 | 0 | 0 | 0 | 0 | 2 | 15 |
| 1979 | 2 | 5 | 2 | 1 | 0 | 3 | 1 | 4 | 2 | 2 | 2 | 1 | 25 |
| 1980 | 1 | 1 | 0 | 2 | 3 | 2 | 3 | 0 | 0 | 2 | 7 | 0 | 21 |
| 1981 | 1 | 6 | 1 | 13 | 3 | 0 | 5 | 1 | 1 | 3 | 1 | 0 | 35 |
| 1982 | 1 | 5 | 2 | 0 | 1 | 16 | 5 | 0 | 0 | 0 | 2 | 10 | 42 |
| 1983 | 0 | 1 | 0 | 0 | 4 | 1 | 0 | 0 | 0 | 0 | 0 | 0 | 6 |
| 1984 | 0 | 1 | 0 | 3 | 3 | 0 | 0 | 0 | 0 | 0 | 0 | 0 | 7 |
| 1985 | 1 | 0 | 0 | 1 | 0 | 0 | 0 | 0 | 0 | 0 | 0 | 0 | 2 |
| 1986 | 0 | 2 | 0 | 0 | 0 | 0 | 0 | 0 | 0 | 0 | 0 | 0 | 2 |
| 1987 | 0 | 0 | 0 | 0 | 0 | 0 | 0 | 0 | 0 | 0 | 0 | 0 | 0 |
| 1988 | 1 | 0 | 0 | 1 | 0 | 4 | 0 | 0 | 0 | 3 | 0 | 4 | 13 |
| 1989 | 7 | 3 | 12 | 1 | 2 | 6 | 1 | 5 | 6 | 5 | 9 | 2 | 59 |
| 1990 | 0 | 0 | 1 | 3 | 8 | 0 | 0 | 1 | 0 | 0 | 0 | 3 | 16 |
| 1991 | 4 | 1 | 20 | 1 | 2 | 6 | 6 | 2 | 2 | 5 | 2 | 3 | 54 |
| 1992 | 1 | 2 | 0 | 0 | 0 | 2 | 1 | 0 | 2 | 1 | 1 | 0 | 10 |
| 1993 | 0 | 0 | 0 | 0 | 0 | 0 | 0 | 0 | 0 | 0 | 0 | 0 | 0 |
| 1994 | 0 | 0 | 0 | 0 | 0 | 0 | 0 | 0 | 0 | 0 | 0 | 0 | 0 |
| 1995 | 0 | 0 | 0 | 0 | 0 | 0 | 0 | 0 | 0 | 0 | 0 | 0 | 0 |
| 1996 | 0 | 0 | 0 | 0 | 0 | 0 | 1 | 0 | 0 | 0 | 0 | 0 | 1 |
| 1997 | 0 | 0 | 0 | 0 | 0 | 0 | 0 | 0 | 0 | 0 | 3 | 0 | 3 |
| 1998 | 0 | 0 | 0 | 2 | 2 | 0 | 0 | 5 | 0 | 0 | 5 | 0 | 14 |
| 1999 | 0 | 0 | 0 | 0 | 0 | 0 | 0 | 2 | 0 | 1 | 1 | 0 | 4 |
| 2000 | 0 | 1 | 3 | 0 | 0 | 4 | 3 | 0 | 1 | 0 | 5 | 0 | 17 |
| 2001 | 0 | 0 | 1 | 8 | 0 | 1 | 0 | 1 | 1 | 4 | 2 | 3 | 21 |
| 2002 | 0 | 0 | 0 | 1 | 1 | 0 | 5 | 4 | 0 | 1 | 0 | 0 | 12 |
| 2003 | 0 | 0 | 2 | 0 | 3 | 4 | 0 | 0 | 0 | 7 | 4 | 0 | 20 |
| 2004 | 0 | 1 | 0 | 0 | 0 | 0 | 6 | 2 | 0 | 1 | 2 | 0 | 12 |
| 2005 | 6 | 0 | 0 | 0 | 0 | 0 | 2 | 0 | 10 | 0 | 0 | 0 | 18 |
| 2006 | 0 | 0 | 0 | 0 | 0 | 0 | 0 | 0 | 0 | 0 | 0 | 4 | 4 |
| 2007 | 0 | 0 | 0 | 0 | 0 | 0 | 0 | 0 | 0 | 0 | 0 | 0 | 0 |
| 2008 | 0 | 0 | 0 | 0 | 0 | 0 | 0 | 0 | 0 | 0 | 0 | 0 | 0 |
| 2009 | 0 | 0 | 0 | 0 | 0 | 0 | 0 | 0 | 0 | 0 | 0 | 0 | 0 |